\pdfoutput=1

\documentclass[11pt]{article}

\usepackage[final]{acl}

\usepackage{times}
\usepackage{latexsym}

\usepackage[T1]{fontenc}

\usepackage[utf8]{inputenc}

\usepackage{microtype}

\usepackage{inconsolata}

\usepackage{graphicx}
\usepackage{amsmath,amsfonts}
\usepackage{algorithmic}
\usepackage{graphicx}
\usepackage{textcomp}
\usepackage{xcolor}
\usepackage{comment}
\usepackage{stfloats} 
\usepackage{graphicx} 
\usepackage{multirow} 
\usepackage{float}    
\usepackage{array}    
\usepackage{subcaption} 
\usepackage{booktabs} 

\def\BibTeX{{\rm B\kern-.05em{\sc i\kern-.025em b}\kern-.08em
    T\kern-.1667em\lower.7ex\hbox{E}\kern-.125emX}}

\usepackage{url} 
\usepackage{float}
\usepackage{enumitem}
\usepackage{xcolor}

\renewcommand{\textcolor}[2]{\begingroup\color{black}#2\endgroup} 

\title{RevPRAG: Revealing Poisoning Attacks in Retrieval-Augmented Generation through LLM Activation Analysis}

\author{
 \textbf{Xue Tan\textsuperscript{1,3}},
 \textbf{Hao Luan\textsuperscript{1,3}},
 \textbf{Mingyu Luo\textsuperscript{1,3}},
 \textbf{Xiaoyan Sun\textsuperscript{2}},
 \textbf{Ping Chen\textsuperscript{3}},
 \textbf{Jun Dai\textsuperscript{2}} \vspace{-0.5em}
\\
 \textsuperscript{1}School of Computer Science, Fudan University, Shanghai, China \vspace{-0.5em}
\\
 \textsuperscript{2}Department of Computer Science, Worcester Polytechnic Institute, MA, USA \vspace{-0.5em}
\\
  \textsuperscript{3}Institute of Big Data, Fudan University, Shanghai, China \vspace{-0.5em}
\\
 \small{
   \textbf{Correspondence:}    \href{mailto:pchen@fudan.edu.cn}{pchen@fudan.edu.cn},
   \href{mailto:xsun7@wpi.edu}{xsun7@wpi.edu},
   \href{mailto:jdai@wpi.edu}{jdai@wpi.edu}
 }
 \setlength{\baselineskip}{0.6\baselineskip}  
}

\begin{document}
\maketitle
\begin{abstract}
Retrieval-Augmented Generation (RAG) enriches the input to LLMs by retrieving information from the relevant knowledge database, enabling them to produce responses that are more accurate and contextually appropriate. 
It is worth noting that the knowledge database, being sourced from publicly available channels such as Wikipedia, inevitably introduces a new attack surface. 
RAG poisoning attack involves injecting malicious texts into the knowledge database, ultimately leading to the generation of the attacker’s target response (also called poisoned response). However, there are currently limited methods available for detecting such poisoning attacks. We aim to bridge the gap in this work by introducing RevPRAG, a flexible and automated detection pipeline that leverages the activations of LLMs for poisoned response detection. Our investigation uncovers distinct patterns in LLMs’ activations when generating poisoned responses versus correct responses. Our results on multiple benchmarks and RAG architectures show our approach can achieve a \emph{98\%} true positive rate, while maintaining a false positive rate close to \emph{1\%}. 
\end{abstract}

\maketitle

\section{Introduction}

\begin{figure*}[t]
    \centering
    \includegraphics[width=0.9\linewidth]{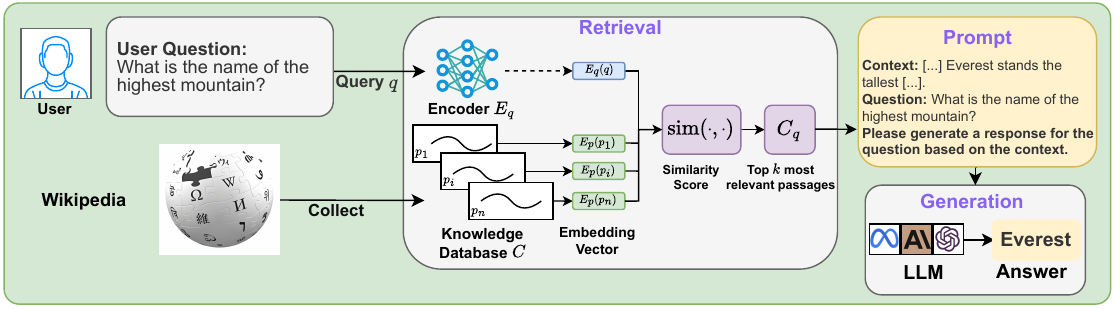}
    \caption{Visualization of RAG.}
    \label{fig:RAG}
\end{figure*}

Retrieval-Augmented Generation (RAG)~\cite{lewis2020retrieval} has emerged as an effective solution that leverages retrievers to incorporate external databases, enriching the knowledge of LLMs and ultimately enabling the generation of up-to-date and accurate responses. RAG comprises three components: \textit{knowledge database}, \textit{retriever}, and \textit{LLM}. Fig.~\ref{fig:RAG} visualizes an example of RAG. The knowledge database consists of a large amount of texts collected from sources such as latest Wikipedia entries~\cite{thakur2021beir}, new articles~\cite{soboroff2018trec} and financial documents~\cite{loukas2023making}. 
The retriever is primarily responsible for retrieving the texts that are most related to the user's query from the knowledge database. These texts will later be fed to LLM as a part of the prompt to generate responses (e.g., ``\textit{Everest}") for users' queries (e.g., ``\textit{What is the name of the highest mountain}?"). 
Due to RAG’s powerful knowledge integration capabilities, it has demonstrated impressive performance across a range of QA-like knowledge-intensive tasks~\cite{lazaridou2022internet, jeong2024adaptive}. 

RAG poisoning refers to the act of injecting malicious or misleading content into the knowledge database, contaminating the retrieved texts and ultimately leading the LLM to produce the attacker’s desired response (e.g., the target answer could be ``\textit{Fuji}" when the target question is ``\textit{What is the name of the highest mountain}?"). This attack leverages the dependency between LLMs and the knowledge database, transforming the database into a new attack surface to facilitate poisoning.
PoisonedRAG ~\cite{zou2024poisonedrag} demonstrates the feasibility of RAG poisoning by injecting a small amount of maliciously crafted texts into the knowledge database utilized by RAG. 
The rise of such attacks has drawn significant attention to the necessity of designing robust and resilient RAG systems.
For example, 
INSTRUCTRAG~\cite{wei2024instructrag} utilizes LLMs to analyze how to extract correct answers from noisy retrieved documents; RobustRAG~\cite{xiang2024certifiably} introduces multiple LLMs to generate answers from the retrieved texts, and then aggregates the responses. However, the aforementioned defense methods necessitate the integration of additional large models, incurring considerable overheads. Meanwhile, it is difficult to promptly assess whether the current response of RAG is trustworthy or not.

In our work, we shift our focus to leverage the \emph{intrinsic} properties of LLMs for detecting RAG poisoning, rather than relying on external models.
Our view is that if we can accurately determine whether a RAG's response is correct or poisoned, we can effectively thwart RAG poisoning attacks. We attempt to observe LLM's answer generation process to determine whether the response is compromised or not. It is worth noting that our focus is not on detecting malicious inputs to LLMs, as we consider the consequences of malicious responses to be far more detrimental and indicative of an attack. The growing body of research on using activations to explain and control LLM behavior~\cite{ferrando2024primer, he2024llm} provides us inspiration.
Specifically, we empirically analyze the activations of the final token in the input sequence across all layers of the LLM. Our findings demonstrate that the model exhibits distinguishable activation patterns when generating correct versus poisoned responses.
Based on this, we propose a systematic and automated detection pipeline, namely RevPRAG, which consists of three key components: \emph{poisoned data collection}, \emph{LLM activation collection and preprocessing}, and the \emph{detection model design}. 
It is important to note that this detection method will not alter the RAG workflow or weaken its performance, thereby offering superior adversarial robustness compared to methods that rely solely on filtering retrieved texts.

To evaluate our approach, we systematically demonstrate the effectiveness of RevPRAG across various LLM architectures, including GPT2-XL-1.5B, Llama2-7B, Mistral-7B, Llama3-8B, and Llama2-13B. 
RevPRAG performs consistently well, achieving over 98\% true positive rate across different datasets. 

Our contributions can be summarized as follows:
\begin{enumerate}

\item We uncover distinct patterns in LLMs’ activations when RAG generates correct responses versus
poisoned ones.


\item We introduce RevPRAG, a novel and automated pipeline for detecting whether a RAG’s response is poisoned or not. To address emerging RAG poisoning attacks, RevPRAG allows new datasets to be constructed accordingly for training the model, enabling effective detection of new threats.

\item Our model has been empirically validated across various LLM architectures and retrievers, demonstrating over 98\% accuracy on our custom-collected detection dataset.
\end{enumerate}

\section{Background and Related Work}
\subsection{Retrieval Augmented Generation}

RAG comprises three components: \textit{knowledge database, retriever,} and \textit{LLM.} 
As illustrated in Fig.~\ref{fig:RAG}, RAG consists of two main steps: \emph{retrieval} step and \emph{generation} step. In the retrieval step, the retriever acquires the top $k$ most relevant pieces of knowledge for the query $q$. First, we employ two encoders, $E_q$ and $E_p$, which can either be identical or radically different. Encoder $E_q$ is responsible for transforming the user's query $q$ into an embedding vector $E_q(q)$, while encoder $E_p$ is designed to convert all the information $p_i$ in the knowledge database into embedding vectors $E_p(p_i)$. For each $E_p(p_i)$, the similarity with the query $E_q(q)$ is computed using $\text{sim}(E_q(q), E_p(p_i))$, where $\text{sim}(\cdot, \cdot)$ quantifies the similarity between two embedding vectors, such as cosine similarity or the dot product. Finally, the top $k$ most relevant pieces are selected as the external knowledge $\mathcal{C}_q$ for the query $q$. The generation step is to generating a response $\text{LLM}(q, \mathcal{C}_q)$ based on the query $q$ and the relevant information $\mathcal{C}_q$. First, we combine the query $q$ and the external knowledge $\mathcal{C}_q$ using a standard prompt (see Fig.~\ref{fig:prompts} for the complete prompt). Taking advantage of such a prompt, the LLM generates an answer $\text{LLM}(q, \mathcal{C}_q)$ to the query $q$.
Therefore, RAG is a significant accomplishment, as it addresses the limitations of LLMs in acquiring up-to-date and domain-specific information.

\subsection{Retrieval Corruption Attack}
Due to the growing attention on RAG, attacks on RAG have also been widely studied. RAG can improperly generate answers that are severely impacted or compromised once the knowledge database is contaminated~\cite{zou2024poisonedrag, xue2024badrag, jiao2024exploring}. 
Specifically, an attacker can inject a small amount of malicious information onto a website, which is then retrieved by RAG~\cite{greshake2023not}. PoisonedRAG~\cite{zou2024poisonedrag} injects malicious text into the knowledge database, and formalizes the knowledge poisoning attack as an optimization problem, thereby enabling the LLM to generate target responses selected by the attacker. GARAG~\cite{cho2024typos} was introduced to provide low-level perturbations to RAG. 
PRCAP~\cite{zhong2023poisoning} injects adversarial samples into the knowledge database, where these samples are generated by perturbing discrete tokens to enhance their similarity with a set of training queries. These methods have yielded striking attack results, and in our work, we have selected several state-of-the-art attack methods as our base attacks on RAG.

\vspace{-5pt}
\subsection{The Robustness of RAG}
Efforts have been made to develop defenses in response to poisoning attacks and noise-induced disruptions. 
RobustRAG~\cite{xiang2024certifiably} mitigates the impact of poisoned texts through a voting mechanism, while INSTRUCTRAG~\cite{wei2024instructrag} explicitly learns the denoising process to address poisoned and irrelevant information. Other approaches to enhance robustness include prompt design~\cite{cho2023improving, press2023measuring}, plug-in models~\cite{baek2023knowledge}, and specialized models~\cite{yoran2023making, asai2023self}. 
However, these methods may, on one hand, rely on additional LLMs, leading to significant overhead. On the other hand, they primarily focus on defense mechanisms before the LLM generates a response, making it challenging for these existing approaches to detect poisoning attacks in real-time while the LLM is generating the response~\cite{athalye2018obfuscated, bryniarski2021evading, carlini2017adversarial, carlini2023llm, tramer2020adaptive}. 
\textcolor{blue}{LLM Factoscope~\cite{ he2024llm}  is a runtime detection tool that leverages the internal states of LLMs, such as activation maps, output rankings, and top-$k$ probabilities, to identify factual inaccuracies caused by model hallucinations. While Factoscope is effective at detecting hallucinations in general LLMs, it is not designed to address RAG poisoning attacks, which result from manipulations of the external knowledge base rather than internal model errors. Its complex architecture with multiple sub-models makes it less suitable for latency-sensitive RAG applications.
In this work, we present RevPRAG, a method that addresses these gaps by: (1) focusing on RAG-specific poisoning attacks and conducting extensive tests to validate its effectiveness in detecting such attacks (Section~\ref{sec:evaluation}), (2) using a lightweight, activation-based pipeline optimized for real-time detection of whether an RAG response is trustworthy (Section~\ref{sec:effi}), and (3) evaluations show that our performance (Section~\ref{overall}) and efficiency (Section~\ref{sec:effi}) surpass those of Factoscope.}


\section{Preliminary}
\subsection{Threat Model}
\vspace{-5pt}
\textbf{Attacker's goal.}
We assume that the attacker preselects a target question set $Q$, consisting of $q_1, q_2, \cdots, q_n$, and the corresponding target answer set $A$, represented as $a_1, a_2, \cdots, a_n$. The attacker's goal is to compromise the RAG system by contaminating the retrieval texts, thereby manipulating the LLM to generate the target response $a_i$ for each query $q_i$.
For example, the attacker’s target question $q_i$ is ``\textit{What is the name of the highest mountain}?", with the target answer being ``\textit{Fuji}". 

\textbf{Attacker's capabilities.}
We assume that an attacker can inject $m$ poisoned texts $P$ for each target question $q_i$, represented as $p_{i}^{1}, p_{i}^{2},...,p_{i}^{m}$. The attacker does not possess knowledge of the LLM utilized by the RAG, but has white-box access to the RAG retriever. This assumption is reasonable, as many retrievers are openly accessible on platforms like HuggingFace.
The poisoned texts can be integrated into the RAG's knowledge database through two ways: the attacker publishing the malicious content on open platforms like Wikipedia, or utilizing data collection agencies to disseminate the poisoned texts.

\begin{figure}[t]
    \centering
    \includegraphics[width=1.0\linewidth]{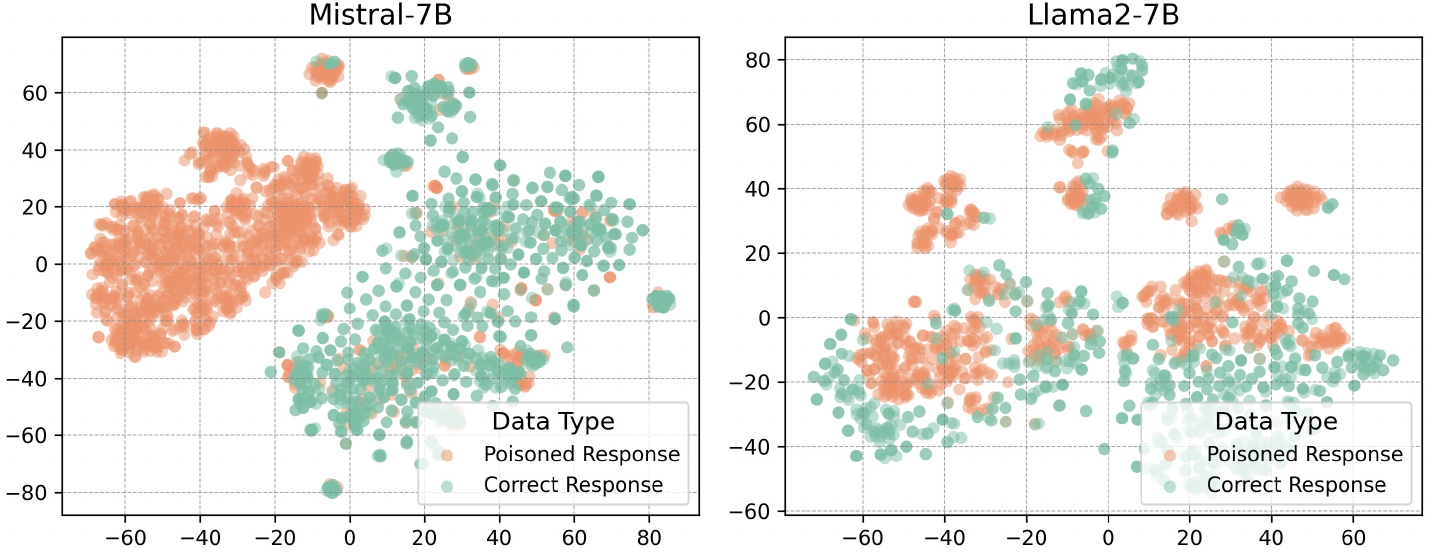}
    \caption{t-SNE visualizations of activations for correct and poisoned responses.}
    \label{fig:vision}
\end{figure}

\subsection{Rationale}
The activations of LLMs represent input data at varying layers of abstraction, enabling the model to progressively extract high-level semantic information from low-level features. The extensive information encapsulated in these activations comprehensively reflects the entire decision-making process of the LLM. The activations has been applied to factual verification of the output content ~\cite{he2024llm} and detection of task drift~\cite{abdelnabi2024you}. Due to the fact that LLM produces different activations when generating varying responses, we hypothesize that LLM will also exhibit distinct activations when generating poisoned responses compared to correct ones.
Fig.~\ref{fig:vision} presents the visualizations of activations for correct and poisoned responses using t-SNE (t-Distributed Stochastic Neighbor Embedding). 
It visualizes the mean activations across all layers for two LLMs, Mistral-7B and Llama2-7B, on the Natural Questions dataset. 
This clearly demonstrates the distinguishability between the two types of responses, to some extent, supports our conjecture. 


\begin{figure*}[b]
    \centering
    \includegraphics[width=0.9\linewidth]{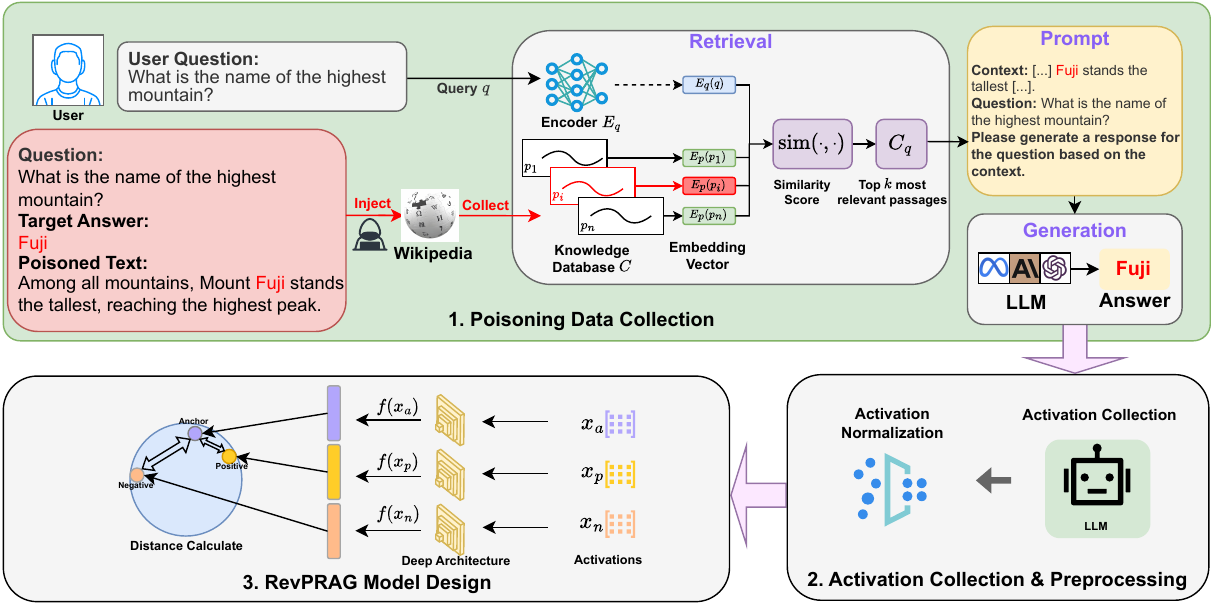}
    \caption{The workflow of RevPRAG.}
    \label{fig:workflow}
\end{figure*}

\section{Methodology}

\subsection{Approach Overview}

\begin{figure}[t]
    \centering
    \includegraphics[width=1.0\linewidth]{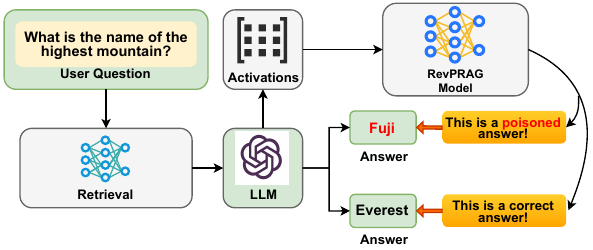}
    \caption{An instance of using RevPRAG.}
    \label{fig:instance}
\end{figure}

As illustrated in Fig.~\ref{fig:workflow}, we introduce RevPRAG, a pipeline designed to leverage LLM activations for detecting knowledge poisoning attacks in RAG systems. 
It contains three major modules: \textit{poisoning data collection, activation collection and preprocessing,} and \textit{RevPRAG detection model design}. 
Fig.~\ref{fig:instance} demonstrates a practical application of RevPRAG for verifying the poisoning status of LLM outputs. Given a user prompt such as “What is the name of the highest mountain?”, the LLM will provide a response. Meanwhile the activations generated by the LLM will be collected and analyzed in RevPRAG. If the model classify the activations as poisoned behavior, it will flag the corresponding response (such as "Fuji") as a poisoned response. Otherwise, it will confirm the response (e.g. "Everest") as the correct answer. 



\subsection{Poisoning Data Collection}
Our method seeks to extract the LLM's activations that capture the model's generation of a specific poisoned response triggered by receiving poisoned texts at a given point in time. Therefore, we first need to implement poisoning attacks on RAG that can mislead the LLM into generating target poisoned responses.
There are three components in RAG: \textit{knowledge database}, \textit{retriever}, and \textit{LLM}. In order to successfully carry out a poisoning attack on RAG and compel the LLM to generate the targeted poisoned response, the initial step is to craft a sufficient amount of poisoned texts and inject them into the knowledge database.
In this paper, in order to create effective poisoned texts for our primary focus on detecting poisoning attacks, we employ three state-of-the-art strategies (i.e., PoisonedRAG~\cite{zou2024poisonedrag}, GARAG~\cite{cho2024typos}, and PAPRAG~\cite{zhong2023poisoning}) for generating poisoned texts and increasing the similarity between the poisoned texts and the queries, to raise the likelihood that the poisoned texts would be selected by the retriever. \textcolor{blue}{A detailed introduction of these methods can be found in Section~\ref{sec:poi}.
The retrieved texts and the question are combined into a new prompt, following the format in~\cite{zou2024poisonedrag} (see Fig.~\ref{fig:prompts} in Section~\ref{sec:prompt}), for LLM answer generation.}

\subsection{Activation Collection and Processing}
\textcolor{blue}{For an LLM input sequence $X=\left ( t_{1},t_{2}, \cdots,t_{n}\right ) $, we extract the activations $Act_{n}$ for the last token $x_{n}$ in the input across all layers in the LLM as a summary of the context. 
The activations $Act_{n}$ contain the inner representations of the LLM's knowledge related to the input. When the LLM generates a response based on a question, it traverses through all layers, retrieving knowledge relevant to the input to produce an answer~\cite{meng2022mass}. 
We collect two types of activations: correct activations (labeled as 1), obtained when the LLM retrieves accurate content and generates the correct response; and poisoned activations (labeled as 0), obtained when the LLM retrieves poisoned content and produces the attacker’s target response.}


\textcolor{blue}{We introduce normalization of the activations for effective integration into the training process. We calculate the mean $\mu$ and standard deviation $\sigma$ of the activations across all instances in the dataset. Then, we use the obtained $\mu$ and $\sigma$ to normalize the activations with the formula: }
\begin{equation}
   Act^{nor}_{n} =\left ( Act_{n}-\mu  \right ) /\sigma.  
\end{equation}


\subsection{RevPRAG Model Design}
\textcolor{blue}{After collecting and preprocessing the activation dataset, we partition it into a training set $D_{train}$, a test set $D_{test}$, and a support set $S$ to facilitate the construction and evaluation of the probe model. Drawing inspiration from few-shot learning and Siamese networks, the proposed RevPRAG model is designed to effectively distinguish between clean and poisoned responses, while demonstrating strong generalization capabilities even under limited data conditions. To efficiently capture both intra-layer and inter-layer relationships within the LLM, we employ Convolutional Neural Networks (CNNs) based on the ResNet18 architecture~\cite{he2016deep}. Additionally, we adopt a triplet network structure, in which three subnetworks with shared architecture and weights are used to learn task embeddings, as illustrated in Fig.~\ref{fig:workflow}.}

\textcolor{blue}{During training, we employ the triplet margin loss~\cite{schroff2015facenet}, a commonly used approach for tasks where it is difficult to distinguish similar instances. The training data is randomly divided into triplets consisting of an anchor instance $x_{a}$, a positive instance $x_{p}$, and a negative instance $x_{n}$, where the anchor and positive belong to the same class, while the anchor and negative come from different classes.
The triplet margin loss function is formally defined as: }

\begin{align}
L = \max\big(&\text{Dist}(x_{a}, x_{p}) - \text{Dist}(x_{a}, x_{n}) \nonumber \\
            &+ \textit{\text{margin}},\ 0\big),
\end{align}
\textcolor{blue}{where $\text{Dist}(\cdot,\cdot)$ denotes a distance metric (typically the Euclidean distance), and \textit{margin} is a positive constant. The training objective is to encourage the RevPRAG embedding model to output closer embedding vectors for any $x_a$ and $x_p$, but farther for any $x_a$ and $x_n$.}

\textcolor{blue}{At test time, given a test sample $x_{t}$, we compute the distance between its embedding and the embedding of the support sample $x_{s}, x_{s}\in S$. The support set $S$ refers to a dataset comprising labeled data, denoted as $\left \{ x_{s_1},...,x_{s_n} \right \} $, and corresponding labels are $\left \{ T_{x_{s_1}},...,T_{x_{s_n}} \right \} $. It provides a reference for comparison and classification of new, unseen test data. The main purpose of the support set is to help determine labels for the test data. The label of the test data $x_{t}$ will be determined according to the label of the support sample $x_{s}$ that is closest to it. That is, $x_{t}$ is assigned the label of $x_{s}$, meaning
$T_{x_t}=T_{x_{s}},$ where $x_s=argmin_i Dist(x_t, x_{s_i})$. Here, $x_{s}$ is the nearest support data to the test data $x_{t}$.}


\newcolumntype{C}[1]{>{\centering\arraybackslash}m{#1}}


\begin{table*}[b]
\caption{ RevPRAG achieved high TPRs and low FPRs on three datasets for RAG with five different LLMs.}
\centering
\scalebox{0.65}{ 
\begin{tabular}{C{2.8cm}C{2cm}C{2cm}C{2cm}C{2cm}C{2cm}C{2cm}}
    \toprule
    \multirow{2}{*}{\Large \textbf{Dataset}} & \multirow{2}{*}{\Large \textbf{Metrics}} & \multicolumn{5}{c}{\Large \textbf{LLMs of RAG}} \\ \cline{3-7}
     &  & \normalsize \textbf{GPT2-XL 1.5B} & \normalsize \textbf{Llama2-7B} & \normalsize \textbf{Mistral-7B} & \normalsize \textbf{Llama3-8B} & \normalsize \textbf{Llama2-13B} \\ \hline
    \multirow{2}{*}{\Large NQ} & \Large TPR &\large {0.982}&\large {0.994}&\large {0.985}&\large {0.986} &\large {0.989} \\ \cline{2-7}
     & \Large FPR &\large {0.006}&\large {0.006}&\large {0.019}&\large {0.009} &\large {0.019}\\ \hline
    \multirow{2}{*}{\Large HotpotQA} & \Large TPR &\large {0.972}&\large {0.985}&\large {0.977}&\large {0.973} &\large {0.970} \\ \cline{2-7}
     & \Large FPR &\large {0.016}&\large {0.061}&\large {0.022}&\large {0.017} &\large {0.070} \\ \hline
    \multirow{2}{*}{\Large MS-MARCO} & \Large TPR &\large {0.988}&\large {0.989}&\large {0.999}&\large {0.978} &\large {0.993}  \\ \cline{2-7}
     & \Large FPR &\large {0.007}&\large {0.012}&\large {0.001}&\large {0.011} &\large {0.025}  \\ 
     \bottomrule
\end{tabular}%
}
\label{tab:different_llms}
\end{table*}

\section{Evaluation}
\label{sec:evaluation}
\subsection{Experimental Setup}
\textbf{RAG Setup.} RAG comprises three key components: \textit{knowledge database, retriever, }and \textit{LLM}.  The setup is shown below:

$\bullet$ \textbf{Knowledge Database:} We leverage three representative benchmark question-answering datasets in our evaluation: Natural Questions (NQ)~\cite{kwiatkowski2019natural}, HotpotQA~\cite{yang2018hotpotqa}, MS-MARCO~\cite{bajaj2016ms}. 
Please note that RevPRAG can be expanded to cover poisoning attacks towards any other datasets used for RAG systems, not limited to the datasets used in this paper. \textcolor{blue}{The detailed usage instructions for the dataset are provided in Section~\ref{sec:train}.}

$\bullet$ \textbf{Retriever:} In our experiments, we evaluate four state-of-the-art dense retrieval models: Contriever~\cite{izacard2021unsupervised} (pre-trained), Contriever-ms (fine-tuned on MS-MARCO)~\cite{izacard2021unsupervised}, DPR-mul~\cite{karpukhin2020dense} (trained on multiple datasets), and ANCE~\cite{xiong2020approximate} (trained on MS-MARCO). 

$\bullet$ \textbf{LLM:} Our experiments are conducted on several popular LLMs, each with distinct architectures and characteristics, including GPT2-XL 1.5B~\cite{radford2019language}, Llama2-7B~\cite{touvron2023llama}, Llama2-13B,  Mistral-7B~\cite{jiang2023mistral}, and Llama3-8B. 

\textcolor{blue}{Unless otherwise specified, we adopt the following default settings: HotpotQA as the knowledge base, Contriever as the retriever, GPT2-XL 1.5B as the LLM, and 100 support samples. Moreover, we use the dot product between the embedding vectors of a question and a text to measure their similarity.  Poisoned texts are generated following PoisonedRAG~\cite{zou2024poisonedrag}. Consistent with prior work~\cite{lewis2020retrieval}, we retrieve the 5 most similar texts from the knowledge database to serve as context for a given question.}

\noindent \textbf{Baselines.} 
We compared RevPRAG with five existing methods, and although they were not specifically designed for detecting RAG poisoning attacks, we investigated their potential applications in this domain. CoS~\cite{li2024chain} is a black-box approach that guides the LLM to generate detailed reasoning steps for the input, subsequently scrutinizing the reasoning process to ensure consistency with the final answer. MDP~\cite{xi2024defending} is a white-box method that exploits the disparity in masking sensitivity between poisoned and clean samples. \textcolor{blue}{LLM Factoscope~\cite{ he2024llm} leverages the internal states of LLMs to detect hallucinations, and we investigate its use for identifying poisoning attacks in RAG systems. Both RoBERTa~\cite{pan2023risk} and Discern~\cite{hong2024so} employ an additional discriminator to distinguish whether the content retrieved by RAG consists of accurate documents or those that contradict factual information.}

\noindent \textbf{Evaluation Metrics.} 


$\bullet$ \textbf{The True Positive Rate (TPR)}, which measures the proportion of effectively poisoned responses that are successfully detected. A higher TPR signifies better detection performance for poisoned responses, with a correspondingly lower rate of missed detections (i.e., lower false negative rate).

$\bullet$ \textbf{The False Positive Rate (FPR)}, which quantifies the proportion of correct responses that are misclassified as being caused by poisoning attacks. A lower FPR indicates fewer false positives for correct answers, minimizing disruption to the normal operation of RAG. Our goal is to detect poisoned responses as effectively as possible while minimizing the impact on RAG’s normal functionality, which is why we have selected these two metrics.

\subsection{Overall Results}\label{overall}

\textbf{RevPRAG achieves high TPRs and low FPRs.} Table~\ref{tab:different_llms} shows the TPRs and FPRs of RevPRAG on three datasets. We have the following observations from the experimental results. First, RevPRAG achieved high TPRs consistently on different datasets and LLMs when injecting five poisoned texts into the knowledge database. For instance, RevPRAG achieved 98.5\% (on NQ), 97.7\% (on HotpotQA), and 99.9\% (on MS-MARCO) TPRs for RAG with Mistral-7B. Our experimental results show that assessing whether the output of a RAG system is correct or poisoned based on the activations of LLMs is both highly feasible and reliable (i.e., capable of achieving exceptional accuracy). Second, RevPRAG achieves low FPRs under different settings, e.g., close to 1\% in nearly all cases. This result indicates that our approach not only maximizes the detection of poisoned responses but also maintains a low false positive rate, significantly reducing the risk of misclassifying correct answers as poisoned. 




\begin{table}[t]
\caption{ RevPRAG achieved high TPRs and low FPRs on HotpotQA for RAG with four different retrievers.}
\centering
\resizebox{\columnwidth}{!}{ 
\begin{tabular}{C{2.8cm}C{2cm}C{2cm}C{2cm}C{2cm}}
    \toprule
    \multirow{2}{*}{\Large \textbf{Attack}} & \multirow{2}{*}{\Large \textbf{Metrics}} & \multicolumn{3}{c}{\Large \textbf{LLMs of RAG}} \\ \cline{3-5}
     &  & \normalsize \textbf{GPT2-XL 1.5B} & \normalsize \textbf{Llama2-7B} & \normalsize \textbf{Mistral-7B} \\ \hline
    \multirow{2}{*}{\Large Contriever} & \Large TPR &\large {0.972}&\large {0.985}&\large {0.977}\\ \cline{2-5}
     & \Large FPR &\large {0.016}&\large {0.061}&\large {0.022}\\ \hline
     \multirow{2}{*}{\Large Contriever-ms} & \Large TPR &\large {0.987}&\large {0.983}&\large {0.998}\\ \cline{2-5}
     & \Large FPR &\large {0.057}&\large {0.018}&\large {0.012}\\ \hline
    \multirow{2}{*}{\Large DPR-mul} & \Large TPR &\large {0.979} &\large {0.966} &\large {0.999}\\ \cline{2-5}
     & \Large FPR &\large {0.035} &\large {0.075} &\large {0.001}\\ \hline
    \multirow{2}{*}{\Large ANCE} & \Large TPR &\large {0.978} &\large {0.981} &\large {0.993}\\ \cline{2-5}
     & \Large FPR &\large {0.042} &\large {0.028} &\large {0.023}\\ 
     \bottomrule
\end{tabular}%
}
\label{tab:contrievers}
\end{table}

\begin{table*}[b]
\caption{ \textcolor{blue}{RevPRAG outperforms baselines.}}
\centering
\scalebox{0.65}{ 
\begin{tabular}{C{2.8cm}C{2cm}C{2.2cm}C{2.2cm}C{2.2cm}C{2.2cm}C{2.2cm}C{2cm}}
    \toprule
    \multirow{2}{*}{\Large \textbf{Dataset}} & \multirow{2}{*}{\Large \textbf{Metrics}} & \multicolumn{6}{c}{\Large \textbf{Baselines and Our Method}} \\ \cline{3-8}
     &  & \normalsize \textbf{CoS~\cite{li2024chain}} & \normalsize \textbf{MDP~\cite{xi2024defending}} & \normalsize \textbf{LLM Factoscope~\cite{ he2024llm}} & \normalsize \textbf{RoBERTa~\cite{pan2023risk}} & \normalsize \textbf{Discern~\cite{hong2024so}} & \normalsize \textbf{Ours} \\ \hline
    \multirow{2}{*}{\Large NQ} & \Large TPR &\large {0.488} &\large {0.946} &\large {0.949} &\large {0.977} &\large {0.810} &\large {\textbf{0.986}} \\ \cline{2-8}
     & \Large FPR &\large {0.146} &\large {0.108} &\large {0.033} &\large {0.063} &\large {0.112} &\large {\textbf{0.009}} \\ \hline
    \multirow{2}{*}{\Large HotpotQA} & \Large TPR &\large {0.194} &\large {0.886} &\large {0.939} &\large {0.956} &\large {0.817} &\large {\textbf{0.973}} \\ \cline{2-8}
     & \Large FPR &\large {0.250} &\large {0.372} &\large {0.021} &\large {0.018} &\large {0.101} &\large {\textbf{0.017}} \\ \hline
    \multirow{2}{*}{\Large MS-MARCO} & \Large TPR &\large {0.771} &\large {\textbf{0.986}} &\large {0.945} &\large {0.946} &\large {0.795} &\large {0.978} \\ \cline{2-8}
     & \Large FPR &\large {0.027} &\large {0.181} &\large {0.028} &\large {0.070} &\large {0.101} &\large {\textbf{0.011}} \\ 
     \bottomrule
\end{tabular}%
}
\label{tab:baselines}
\end{table*}

We also conduct experiments on different retrievers. Table~\ref{tab:contrievers} 
shows that our approach consistently achieved high TPRs and low FPRs across RAG with various retrievers and LLMs. For instance, RevPRAG achieves 97.2\% (with Contriever), 98.7\% (with Contriever-ms), 97.9\% (with DPR-mul), 97.8\% (with ANCE) TPRs alongside 1.6\% (with Contriever), 5.7\% (with Contriever-ms), 3.5\% (with DPR-mul), and 4.2\% (with ANCE) FPRs for RAG when using GPT2-XL 1.5B. 

\textbf{RevPRAG outperforms baselines.} \textcolor{blue}{Table~\ref{tab:baselines} compares RevPRAG with baselines for RAG using Llama3-8B under the default settings. The overall results demonstrate the superiority of our approach. Meanwhile, several key observations can be drawn from the comparison. First, the limited effectiveness of CoS~\cite{li2024chain} may stem from its design focus on detecting backdoor attacks in LLMs via trigger-to-output shortcuts, which differs from RAG's attack surface involving poisoned knowledge base entries. Second, MDP~\cite{xi2024defending} achieves good TPRs, but it also exhibits relatively high FPRs, reaching as much as 37.2\%. LLM Factoscope~\cite{ he2024llm} leverages multiple internal states of LLMs, relying on layer-wise consistency for effective hallucination detection. However, it may not be suitable for targeted attacks like poisoning, and the use of diverse state data increases computational overhead and discriminator model complexity (Section~\ref{sec:effi}). Input-based methods such as MDP~\cite{xi2024defending}, RoBERTa~\cite{pan2023risk}, and Discern~\cite{hong2024so} aim to detect whether the \emph{input} is poisoned. In contrast, our method focuses on determining whether the responses generated by RAG are correct or poisoned, as response correctness offers a more robust signal of poisoning attacks. 
}

\subsection{Ablation Study}

\textbf{Different methods for generating poisoned texts.}
To ensure the effectiveness of the evaluation, we employ three different methods introduced by PoisonedRAG, GARAG, and PRCAP to generate the poisoned texts. The experimental results in Table~\ref{tab:texts} show that RevPRAG consistently achieves high TPRs and low FPRs when confronted with poisoned texts generated by different strategies. For instance, RevPRAG achieved 97.2\% (with GPT2-XL 1.5B), 98.5\% (with Llama2-7B), and 97.7\% (with Mistral-7B) TPRs for poisoned texts generated with PoisonedRAG. 
\begin{table}[t]
\caption{ The TPRs and FPRs of RevPRAG for different poisoned text generation methods on HotpotQA.}
\centering
\resizebox{\columnwidth}{!}{ 
\begin{tabular}{C{2.8cm}C{2cm}C{2cm}C{2cm}C{2cm}}
    \toprule
    \multirow{2}{*}{\Large \textbf{Attack}} & \multirow{2}{*}{\Large \textbf{Metrics}} & \multicolumn{3}{c}{\Large \textbf{LLMs of RAG}} \\ \cline{3-5}
     &  & \normalsize \textbf{GPT2-XL 1.5B} & \normalsize \textbf{Llama2-7B} & \normalsize \textbf{Mistral-7B} \\ \hline
    \multirow{2}{*}{\Large PoisonedRAG} & \Large TPR &\large {0.972}&\large {0.985}&\large {0.977}\\ \cline{2-5}
     & \Large FPR &\large {0.016}&\large {0.061}&\large {0.022}\\ \hline
    \multirow{2}{*}{\Large GARAG} & \Large TPR &\large {0.961} &\large {0.976} &\large {0.974}\\ \cline{2-5}
     & \Large FPR &\large {0.025} &\large {0.046} &\large {0.026}\\ \hline
    \multirow{2}{*}{\Large PRCAP} & \Large TPR &\large {0.966} &\large {0.986} &\large {0.965}\\ \cline{2-5}
     & \Large FPR &\large {0.012} &\large {0.061} &\large {0.022}\\ 
     \bottomrule
\end{tabular}%
}
\label{tab:texts}
\end{table}

\begin{table}[H]
\caption{ The TPRs and FPRs of RevPRAG for different quantities of injected poisoned text on HotpotQA (total retrieved texts: five).}
\centering
\resizebox{\columnwidth}{!}{ 
\begin{tabular}{C{2.5cm}C{2cm}C{2cm}C{2cm}C{2cm}}
    \toprule
    \multirow{2}{*}{\Large \textbf{Quantity}} & \multirow{2}{*}{\Large \textbf{Metrics}} & \multicolumn{3}{c}{\Large \textbf{LLMs of RAG}} \\ \cline{3-5}
     &  & \normalsize \textbf{GPT2-XL 1.5B} & \normalsize \textbf{Llama2-7B} & \normalsize \textbf{Mistral-7B} \\ \hline
    \multirow{2}{*}{\Large five} & \Large TPR &\large {0.972}&\large {0.985}&\large {0.977}\\ \cline{2-5}
     & \Large FPR &\large {0.016}&\large {0.061}&\large {0.022}\\ \hline
    \multirow{2}{*}{\Large four} & \Large TPR &\large {0.976}&\large {0.977}&\large {0.986}\\ \cline{2-5}
     & \Large FPR &\large {0.034}&\large {0.047}&\large {0.033}\\ \hline
    \multirow{2}{*}{\Large three} & \Large TPR &\large {0.963}&\large {0.986}&\large {0.995}\\ \cline{2-5}
     & \Large FPR &\large {0.011}&\large {0.043}&\large {0.004}\\ \hline
    \multirow{2}{*}{\Large two} & \Large TPR &\large {0.971}&\large {0.995}&\large {0.991}\\ \cline{2-5}
     & \Large FPR &\large {0.011}&\large {0.047}&\large {0.005}\\ \hline
    \multirow{2}{*}{\Large one} & \Large TPR &\large {0.970}&\large {0.988}&\large {0.989}\\ \cline{2-5}
     & \Large FPR &\large {0.049}&\large {0.031}&\large {0.022}\\
     \bottomrule
\end{tabular}%
}
\label{tab:number}
\end{table}

\textbf{Quantity of injected poisoned texts.} Table~\ref{tab:number} illustrates the impact of varying quantities of poisoned text on the detection performance of RevPRAG. 
The more poisoned texts are injected, the higher the likelihood of retrieving them for RAG processing. 
From the experimental results, we observe that even with varying amounts of injected poisoned text, RevPRAG consistently achieves high TPRs and low FPRs. For example, when the total number of retrieved texts is five and the injected quantity is two, RevPRAG achieves a 99.5\% TPR and a 4.7\% FPR for RAG with Llama2-7B. 
The reason for this phenomenon is that the similarity between the retrieved poisoned texts and the query is higher than that of clean texts. Consequently, the LLM generates responses based on the content of the poisoned texts.

\begin{figure}[h]
    \centering
    \includegraphics[width=1.0\linewidth]{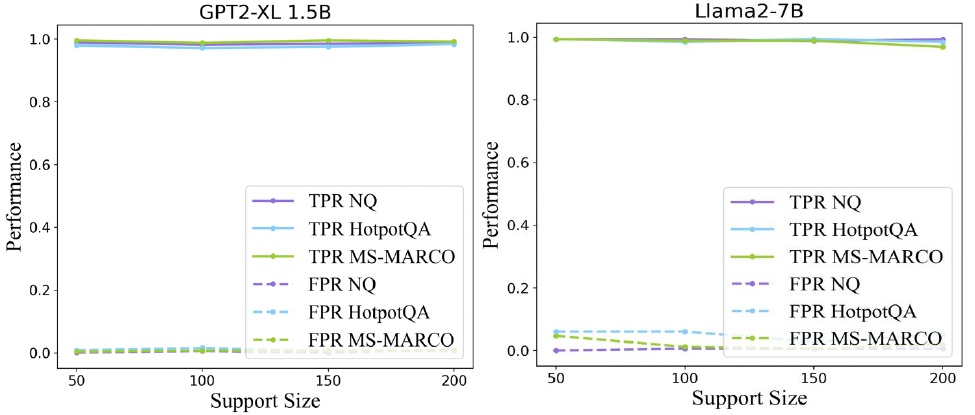}
    \caption{Effects of support set size.}
    \label{fig:support}
\end{figure}

\textbf{Effects of different support set size.} In RevPRAG, support data provides essential labeled and task-specific information, facilitating effective reasoning and learning under limited data conditions. We experiment with various support set sizes ranging from 50 to 250 to examine their effect on the performance of RevPRAG. 
The results in Fig.~\ref{fig:support} indicate that varying the support size does not significantly impact the model’s detection performance. 

\subsection{RevPRAG’s Performance on Complex Open-Ended Questions} In this section, we conducted a series of experiments to evaluate the performance of RevPRAG on complex, open-ended questions (e.g., “\textit{how to make relationship last?}”). These questions present unique challenges due to their diverse and unstructured nature, in contrast to straightforward, closed-ended questions (e.g., “\textit{What is the name of the highest mountain?}”). In our experiments, the NQ, HotpotQA, and MS-MARCO datasets primarily consist of close-ended questions. As a result, the majority of our previous experiments focused on close-ended problems, which was our default experimental setting. In this study, we utilized the advanced GPT-4o to filter and extract 3,000 open-ended questions from the HotpotQA and MS-MARCO datasets for training and testing the model. For open-ended questions, cosine similarity is employed to evaluate whether the LLM’s response aligns with the attacker’s target response. If the similarity surpasses a predefined threshold, it is considered indicative of a successful attack.

\newcolumntype{C}[1]{>{\centering\arraybackslash}m{#1}}

\begin{table}[htbp]
\caption{ RevPRAG achieved high TPRs and low FPRs on the open-ended questions from HotpotQA and MS-MARCO datasets.}
\centering
\resizebox{\columnwidth}{!}{ 
\begin{tabular}{C{3.5cm}C{2cm}C{2cm}C{2cm}C{2cm}C{2cm}}
    \hline
    \multirow{2}{*}{\Large \textbf{Dataset}} & \multirow{2}{*}{\Large \textbf{Metrics}} & \multicolumn{4}{c}{\Large \textbf{LLMs of RAG}} \\ \cline{3-6}
     &  & \normalsize \textbf{GPT2-XL 1.5B} & \normalsize \textbf{Llama2-7B} & \normalsize \textbf{Mistral-7B} & \normalsize \textbf{Llama3-8B} \\ \hline
    \multirow{2}{*}{\Large HotpotQA} & \Large TPR &\large {0.982}&\large {0.995}&\large {0.991}&\large {0.982} \\ \cline{2-6}
     & \Large FPR &\large {0.033}&\large {0.029}&\large {0.008}&\large {0.007} \\ \hline
    \multirow{2}{*}{\Large MS-MARCO} & \Large TPR &\large {0.988}&\large {0.989}&\large {0.990}&\large {0.983}  \\ \cline{2-6}
     & \Large FPR &\large {0.009}&\large {0.009}&\large {0.001}&\large {0.017} \\ \hline
     
\end{tabular}%
}
\label{tab:open-ended}
\end{table}

The experimental results are shown in Table~\ref{tab:open-ended}. We can observe that RevPRAG demonstrates excellent detection performance even on complex open-ended questions. For example, RevPRAG achieved TPRs of 99.1\% on HotpotQA and 99.0\% on MS-MARCO, alongside FPRs of 0.8\% on HotpotQA and 0.1\% on MS-MARCO for RAG utilizing the Mistral-7B model. 

\subsection{Effect of Real-world Natural Text Noise on Detection Performance}
\label{sec:noise}

To better approximate real-world scenarios, we injected natural textual noise (e.g., spelling and grammatical errors) into clean texts using the GPT-4o model. Noise was introduced in a controlled manner to preserve semantics while simulating realistic imperfections. Specifically, 50\% of the clean texts were modified, with noise accounting for 10\% of the total word count in each case. Notably, noise injection was applied only to clean texts, as poisoned texts are typically well-formed to maximize attack effectiveness and thus unlikely to contain such artifacts.  

As shown in Table~\ref{tab:attack}, the experimental results indicate that injecting a certain amount of natural noise into clean texts does not significantly affect the detection performance of our proposed method. This further demonstrates the robustness of our approach. We attribute this to two main factors. First, the textual perturbations introduced by poisoning attacks are fundamentally different in nature from natural noise, and this distinction is clearly reflected in the activation vectors of the LLM. Second, LLMs are inherently robust to natural noise, which does not substantially interfere with their ability to generate correct responses.


\newcolumntype{C}[1]{>{\centering\arraybackslash}m{#1}}
\begin{table}[htbp]
\caption{ \textcolor{blue}{Performance of RevPRAG on noisy datasets.}}
\centering
\resizebox{\columnwidth}{!}{ 
\begin{tabular}{C{3.2cm}C{1.8cm}C{2cm}C{2cm}}
    \toprule
    {\textbf{Dataset}} & { \textbf{Metrics}} & \textbf{GPT2-XL 1.5B} & \textbf{Llama3-8B} \\ \hline
    \multirow{2}{*}{ Noisy\_NQ} &  TPR & {0.980}& {0.977} \\ \cline{2-4}
     &  FPR & {0.034}& {0.012} \\ \hline
    \multirow{2}{*}{ Noisy\_HotpotQA} &  TPR & {0.969}& {0.974}  \\ \cline{2-4}
     &  FPR & {0.026}& {0.011} \\ \hline
    \multirow{2}{*}{ Noisy\_MS-MARCO} &  TPR & {0.989}& {0.977}  \\ \cline{2-4}
     &  FPR & {0.013}& {0.018}   \\ 
     \bottomrule
\end{tabular}%
}
\label{tab:attack}
\end{table}

\subsection{Efficiency}
\label{sec:effi}
\textcolor{blue}{Table~\ref{tab:effi} compares the time overhead between LLM Factoscope~\cite{ he2024llm} and RevPRAG when the LLM in RAG is Llama3-8B, including the average training time per epoch and the average inference time per test sample. This experiment was conducted using 1,000 training samples and 500 test samples, with poisoned and clean examples each accounting for 50\%. The results demonstrate that RevPRAG, with its task-specific architecture and carefully selected detection metrics, incurs significantly lower computational costs than LLM Factoscope, which integrates multiple sub-models for hallucination detection. Its efficient detection capability makes RevPRAG particularly well-suited for latency-sensitive RAG scenarios, underscoring its practical value.}

\begin{table}[h]
\centering
\caption{ \textcolor{blue}{Comparison of time overhead.}}
\label{tab:effi}
\resizebox{0.5\textwidth}{!}{ 
\begin{tabular}{ccccc}

\toprule
\multirow{2}{*}{\textbf{Dataset}} & \multicolumn{2}{c}{\textbf{Training Time per Epoch}} & \multicolumn{2}{c}{\textbf{Inference Time per Sample}} \\
\cline{2-5}
&LLM factoscope & RevPRAG & LLM factoscope & RevPRAG \\
 
\hline
NQ & 91.61s & \textbf{19.31s} & 0.0051s & \textbf{0.0021s} \\
\hline
HotpotQA & 101.25s & \textbf{23.69s} & 0.0066s & \textbf{0.0023s}  \\
\hline
MS-MARCO & 94.47s & \textbf{20.72s} & 0.0058s & \textbf{0.0022s}  \\
\bottomrule
\end{tabular}}

\end{table}

\vspace{-3pt}
\section{Conclusion}

In this work, we find that correct and poisoned responses in RAG exhibit distinct differences in LLM activations. Building on this insight, we develop RevPRAG, a detection pipeline that leverages these activations to identify poisoned responses in RAG caused by the injection of malicious texts into the knowledge database.
Our approach demonstrates robust performance across RAGs utilizing five different LLMs and four distinct retrievers on three datasets. Experimental results show that RevPRAG achieves exceptional accuracy, with true positive rates approaching 98\% and false positive rates near 1\%. Ablation studies further validate its effectiveness in detecting poisoned responses across different types and levels of poisoning attacks. Overall, our approach can accurately distinguish between correct and poisoned responses.

\textbf{Limitations.}

Our work has the following limitations:

\begin{itemize}
  \item This work does not propose a specific method for defending against poisoning attacks on RAG. Instead, our focus is on the timely detection of poisoned responses generated by the LLM, aiming to prevent potential harm to users from such attacks. 
  \item Our approach requires accessing the activations of the LLM, which necessitates the LLM being a white-box model. While this may present certain limitations for users, our method can be widely adopted by LLM service providers. Providers can implement our strategy to ensure the reliability of their services and enhance trust with their users. 
  \item Our approach primarily focuses on determining whether the response generated by the RAG is correct or poisoned, without delving into more granular distinctions. The main goal of our study is to protect users from the impact of RAG poisoning attacks, while more detailed classifications of RAG responses will be addressed in future work.
\end{itemize}

\textbf{Ethics Statement}

The goal of this work is to detect whether a RAG has generated a poisoned response. All the data used in this study is publicly available, so it does not introduce additional privacy concerns. All source code and software will be made open-source. While the open-source nature of the code may lead to adaptive attacks, we can further enhance our model by incorporating more internal and external information. Overall, we believe our approach can further promote the secure application of RAG.

\bibliography{ref}

\begin{thebibliography}{44}
\providecommand{\natexlab}[1]{#1}

\bibitem[{Abdelnabi et~al.(2024)Abdelnabi, Fay, Cherubin, Salem, Fritz, and Paverd}]{abdelnabi2024you}
Sahar Abdelnabi, Aideen Fay, Giovanni Cherubin, Ahmed Salem, Mario Fritz, and Andrew Paverd. 2024.
\newblock Are you still on track!? catching llm task drift with activations.
\newblock \emph{arXiv preprint arXiv:2406.00799}.

\bibitem[{Asai et~al.(2023)Asai, Wu, Wang, Sil, and Hajishirzi}]{asai2023self}
Akari Asai, Zeqiu Wu, Yizhong Wang, Avirup Sil, and Hannaneh Hajishirzi. 2023.
\newblock Self-rag: Learning to retrieve, generate, and critique through self-reflection.
\newblock \emph{arXiv preprint arXiv:2310.11511}.

\bibitem[{Athalye et~al.(2018)Athalye, Carlini, and Wagner}]{athalye2018obfuscated}
Anish Athalye, Nicholas Carlini, and David Wagner. 2018.
\newblock Obfuscated gradients give a false sense of security: Circumventing defenses to adversarial examples.
\newblock In \emph{International conference on machine learning}, pages 274--283. PMLR.

\bibitem[{Baek et~al.(2023)Baek, Jeong, Kang, Park, and Hwang}]{baek2023knowledge}
Jinheon Baek, Soyeong Jeong, Minki Kang, Jong~C Park, and Sung Hwang. 2023.
\newblock Knowledge-augmented language model verification.
\newblock In \emph{Proceedings of the 2023 Conference on Empirical Methods in Natural Language Processing}, pages 1720--1736.

\bibitem[{Bajaj et~al.(2016)Bajaj, Campos, Craswell, Deng, Gao, Liu, Majumder, McNamara, Mitra, Nguyen et~al.}]{bajaj2016ms}
Payal Bajaj, Daniel Campos, Nick Craswell, Li~Deng, Jianfeng Gao, Xiaodong Liu, Rangan Majumder, Andrew McNamara, Bhaskar Mitra, Tri Nguyen, and 1 others. 2016.
\newblock Ms marco: A human generated machine reading comprehension dataset.
\newblock \emph{arXiv preprint arXiv:1611.09268}.

\bibitem[{Bryniarski et~al.(2021)Bryniarski, Hingun, Pachuca, Wang, and Carlini}]{bryniarski2021evading}
Oliver Bryniarski, Nabeel Hingun, Pedro Pachuca, Vincent Wang, and Nicholas Carlini. 2021.
\newblock Evading adversarial example detection defenses with orthogonal projected gradient descent.
\newblock \emph{arXiv preprint arXiv:2106.15023}.

\bibitem[{Carlini(2023)}]{carlini2023llm}
Nicholas Carlini. 2023.
\newblock A llm assisted exploitation of ai-guardian.
\newblock \emph{arXiv preprint arXiv:2307.15008}.

\bibitem[{Carlini and Wagner(2017)}]{carlini2017adversarial}
Nicholas Carlini and David Wagner. 2017.
\newblock Adversarial examples are not easily detected: Bypassing ten detection methods.
\newblock In \emph{Proceedings of the 10th ACM workshop on artificial intelligence and security}, pages 3--14.

\bibitem[{Cho et~al.(2024)Cho, Jeong, Seo, Hwang, and Park}]{cho2024typos}
Sukmin Cho, Soyeong Jeong, Jeongyeon Seo, Taeho Hwang, and Jong~C Park. 2024.
\newblock Typos that broke the rag's back: Genetic attack on rag pipeline by simulating documents in the wild via low-level perturbations.
\newblock \emph{arXiv preprint arXiv:2404.13948}.

\bibitem[{Cho et~al.(2023)Cho, Seo, Jeong, and Park}]{cho2023improving}
Sukmin Cho, Jeongyeon Seo, Soyeong Jeong, and Jong~C Park. 2023.
\newblock Improving zero-shot reader by reducing distractions from irrelevant documents in open-domain question answering.
\newblock In \emph{Findings of the Association for Computational Linguistics: EMNLP 2023}, pages 3145--3157.

\bibitem[{Ferrando et~al.(2024)Ferrando, Sarti, Bisazza, and Costa-juss{\`a}}]{ferrando2024primer}
Javier Ferrando, Gabriele Sarti, Arianna Bisazza, and Marta~R Costa-juss{\`a}. 2024.
\newblock A primer on the inner workings of transformer-based language models.
\newblock \emph{arXiv preprint arXiv:2405.00208}.

\bibitem[{Greshake et~al.(2023)Greshake, Abdelnabi, Mishra, Endres, Holz, and Fritz}]{greshake2023not}
Kai Greshake, Sahar Abdelnabi, Shailesh Mishra, Christoph Endres, Thorsten Holz, and Mario Fritz. 2023.
\newblock Not what you've signed up for: Compromising real-world llm-integrated applications with indirect prompt injection.
\newblock In \emph{Proceedings of the 16th ACM Workshop on Artificial Intelligence and Security}, pages 79--90.

\bibitem[{He et~al.(2024)He, Gong, Lin, Zhao, Chen et~al.}]{he2024llm}
Jinwen He, Yujia Gong, Zijin Lin, Yue Zhao, Kai Chen, and 1 others. 2024.
\newblock Llm factoscope: Uncovering llms’ factual discernment through measuring inner states.
\newblock In \emph{Findings of the Association for Computational Linguistics ACL 2024}, pages 10218--10230.

\bibitem[{He et~al.(2016)He, Zhang, Ren, and Sun}]{he2016deep}
Kaiming He, Xiangyu Zhang, Shaoqing Ren, and Jian Sun. 2016.
\newblock Deep residual learning for image recognition.
\newblock In \emph{Proceedings of the IEEE conference on computer vision and pattern recognition}, pages 770--778.

\bibitem[{Hong et~al.(2024)Hong, Kim, Kang, Myaeng, and Whang}]{hong2024so}
Giwon Hong, Jeonghwan Kim, Junmo Kang, Sung-Hyon Myaeng, and Joyce Whang. 2024.
\newblock Why so gullible? enhancing the robustness of retrieval-augmented models against counterfactual noise.
\newblock In \emph{Findings of the Association for Computational Linguistics: NAACL 2024}, pages 2474--2495.

\bibitem[{Izacard et~al.(2021)Izacard, Caron, Hosseini, Riedel, Bojanowski, Joulin, and Grave}]{izacard2021unsupervised}
Gautier Izacard, Mathilde Caron, Lucas Hosseini, Sebastian Riedel, Piotr Bojanowski, Armand Joulin, and Edouard Grave. 2021.
\newblock Unsupervised dense information retrieval with contrastive learning.
\newblock \emph{arXiv preprint arXiv:2112.09118}.

\bibitem[{Jeong et~al.(2024)Jeong, Baek, Cho, Hwang, and Park}]{jeong2024adaptive}
Soyeong Jeong, Jinheon Baek, Sukmin Cho, Sung~Ju Hwang, and Jong~C Park. 2024.
\newblock Adaptive-rag: Learning to adapt retrieval-augmented large language models through question complexity.
\newblock In \emph{Proceedings of the 2024 Conference of the North American Chapter of the Association for Computational Linguistics: Human Language Technologies (Volume 1: Long Papers)}, pages 7029--7043.

\bibitem[{Jiang et~al.(2023)Jiang, Sablayrolles, Mensch, Bamford, Chaplot, Casas, Bressand, Lengyel, Lample, Saulnier et~al.}]{jiang2023mistral}
Albert~Q Jiang, Alexandre Sablayrolles, Arthur Mensch, Chris Bamford, Devendra~Singh Chaplot, Diego de~las Casas, Florian Bressand, Gianna Lengyel, Guillaume Lample, Lucile Saulnier, and 1 others. 2023.
\newblock Mistral 7b.
\newblock \emph{arXiv preprint arXiv:2310.06825}.

\bibitem[{Jiao et~al.(2024)Jiao, Xie, Yue, Sato, Wang, Wang, Chen, and Zhu}]{jiao2024exploring}
Ruochen Jiao, Shaoyuan Xie, Justin Yue, Takami Sato, Lixu Wang, Yixuan Wang, Qi~Alfred Chen, and Qi~Zhu. 2024.
\newblock Exploring backdoor attacks against large language model-based decision making.
\newblock \emph{arXiv preprint arXiv:2405.20774}.

\bibitem[{Karpukhin et~al.(2020)Karpukhin, O{\u{g}}uz, Min, Lewis, Wu, Edunov, Chen, and Yih}]{karpukhin2020dense}
Vladimir Karpukhin, Barlas O{\u{g}}uz, Sewon Min, Patrick Lewis, Ledell Wu, Sergey Edunov, Danqi Chen, and Wen~Tau Yih. 2020.
\newblock Dense passage retrieval for open-domain question answering.
\newblock In \emph{2020 Conference on Empirical Methods in Natural Language Processing, EMNLP 2020}, pages 6769--6781.

\bibitem[{Kwiatkowski et~al.(2019)Kwiatkowski, Palomaki, Redfield, Collins, Parikh, Alberti, Epstein, Polosukhin, Devlin, Lee et~al.}]{kwiatkowski2019natural}
Tom Kwiatkowski, Jennimaria Palomaki, Olivia Redfield, Michael Collins, Ankur Parikh, Chris Alberti, Danielle Epstein, Illia Polosukhin, Jacob Devlin, Kenton Lee, and 1 others. 2019.
\newblock Natural questions: a benchmark for question answering research.
\newblock \emph{Transactions of the Association for Computational Linguistics}, 7:453--466.

\bibitem[{Lazaridou et~al.(2022)Lazaridou, Gribovskaya, Stokowiec, and Grigorev}]{lazaridou2022internet}
Angeliki Lazaridou, Elena Gribovskaya, Wojciech Stokowiec, and Nikolai Grigorev. 2022.
\newblock Internet-augmented language models through few-shot prompting for open-domain question answering.
\newblock \emph{arXiv preprint arXiv:2203.05115}.

\bibitem[{Lewis et~al.(2020)Lewis, Perez, Piktus, Petroni, Karpukhin, Goyal, K{\"u}ttler, Lewis, Yih, Rockt{\"a}schel et~al.}]{lewis2020retrieval}
Patrick Lewis, Ethan Perez, Aleksandra Piktus, Fabio Petroni, Vladimir Karpukhin, Naman Goyal, Heinrich K{\"u}ttler, Mike Lewis, Wen-tau Yih, Tim Rockt{\"a}schel, and 1 others. 2020.
\newblock Retrieval-augmented generation for knowledge-intensive nlp tasks.
\newblock \emph{Advances in Neural Information Processing Systems}, 33:9459--9474.

\bibitem[{Li et~al.(2024)Li, Zhang, Lou, Wu, and Wang}]{li2024chain}
Xi~Li, Yusen Zhang, Renze Lou, Chen Wu, and Jiaqi Wang. 2024.
\newblock Chain-of-scrutiny: Detecting backdoor attacks for large language models.
\newblock \emph{arXiv preprint arXiv:2406.05948}.

\bibitem[{Loukas et~al.(2023)Loukas, Stogiannidis, Diamantopoulos, Malakasiotis, and Vassos}]{loukas2023making}
Lefteris Loukas, Ilias Stogiannidis, Odysseas Diamantopoulos, Prodromos Malakasiotis, and Stavros Vassos. 2023.
\newblock Making llms worth every penny: Resource-limited text classification in banking.
\newblock In \emph{Proceedings of the Fourth ACM International Conference on AI in Finance}, pages 392--400.

\bibitem[{Meng et~al.(2023)Meng, Sharma, Andonian, Belinkov, and Bau}]{meng2022mass}
Kevin Meng, Arnab~Sen Sharma, Alex~J Andonian, Yonatan Belinkov, and David Bau. 2023.
\newblock Mass-editing memory in a transformer.
\newblock In \emph{The Eleventh International Conference on Learning Representations}.

\bibitem[{Pan et~al.(2023)Pan, Pan, Chen, Nakov, Kan, and Wang}]{pan2023risk}
Yikang Pan, Liangming Pan, Wenhu Chen, Preslav Nakov, Min-Yen Kan, and William Wang. 2023.
\newblock On the risk of misinformation pollution with large language models.
\newblock In \emph{Findings of the Association for Computational Linguistics: EMNLP 2023}, pages 1389--1403.

\bibitem[{Press et~al.(2023)Press, Zhang, Min, Schmidt, Smith, and Lewis}]{press2023measuring}
Ofir Press, Muru Zhang, Sewon Min, Ludwig Schmidt, Noah~A Smith, and Mike Lewis. 2023.
\newblock Measuring and narrowing the compositionality gap in language models.
\newblock In \emph{Findings of the Association for Computational Linguistics: EMNLP 2023}, pages 5687--5711.

\bibitem[{Radford et~al.(2019)Radford, Wu, Child, Luan, Amodei, Sutskever et~al.}]{radford2019language}
Alec Radford, Jeffrey Wu, Rewon Child, David Luan, Dario Amodei, Ilya Sutskever, and 1 others. 2019.
\newblock Language models are unsupervised multitask learners.
\newblock \emph{OpenAI blog}, 1(8):9.

\bibitem[{Schroff et~al.(2015)Schroff, Kalenichenko, and Philbin}]{schroff2015facenet}
Florian Schroff, Dmitry Kalenichenko, and James Philbin. 2015.
\newblock Facenet: A unified embedding for face recognition and clustering.
\newblock In \emph{Proceedings of the IEEE conference on computer vision and pattern recognition}, pages 815--823.

\bibitem[{Soboroff et~al.(2018)Soboroff, Huang, and Harman}]{soboroff2018trec}
Ian Soboroff, Shudong Huang, and Donna Harman. 2018.
\newblock Trec 2018 news track overview.
\newblock In \emph{TREC}, volume 409, page 410.

\bibitem[{Thakur et~al.(2021)Thakur, Reimers, R{\"u}ckl{\'e}, Srivastava, and Gurevych}]{thakur2021beir}
Nandan Thakur, Nils Reimers, Andreas R{\"u}ckl{\'e}, Abhishek Srivastava, and Iryna Gurevych. 2021.
\newblock Beir: A heterogenous benchmark for zero-shot evaluation of information retrieval models.
\newblock \emph{arXiv preprint arXiv:2104.08663}.

\bibitem[{Touvron et~al.(2023)Touvron, Lavril, Izacard, Martinet, Lachaux, Lacroix, Rozi{\`e}re, Goyal, Hambro, Azhar et~al.}]{touvron2023llama}
Hugo Touvron, Thibaut Lavril, Gautier Izacard, Xavier Martinet, Marie-Anne Lachaux, Timoth{\'e}e Lacroix, Baptiste Rozi{\`e}re, Naman Goyal, Eric Hambro, Faisal Azhar, and 1 others. 2023.
\newblock Llama: Open and efficient foundation language models.
\newblock \emph{arXiv preprint arXiv:2302.13971}.

\bibitem[{Tramer et~al.(2020)Tramer, Carlini, Brendel, and Madry}]{tramer2020adaptive}
Florian Tramer, Nicholas Carlini, Wieland Brendel, and Aleksander Madry. 2020.
\newblock On adaptive attacks to adversarial example defenses.
\newblock \emph{Advances in neural information processing systems}, 33:1633--1645.

\bibitem[{Wei et~al.(2024)Wei, Chen, and Meng}]{wei2024instructrag}
Zhepei Wei, Wei-Lin Chen, and Yu~Meng. 2024.
\newblock Instructrag: Instructing retrieval-augmented generation with explicit denoising.
\newblock \emph{arXiv preprint arXiv:2406.13629}.

\bibitem[{Xi et~al.(2024)Xi, Du, Li, Pang, Ji, Chen, Ma, and Wang}]{xi2024defending}
Zhaohan Xi, Tianyu Du, Changjiang Li, Ren Pang, Shouling Ji, Jinghui Chen, Fenglong Ma, and Ting Wang. 2024.
\newblock Defending pre-trained language models as few-shot learners against backdoor attacks.
\newblock \emph{Advances in Neural Information Processing Systems}, 36.

\bibitem[{Xiang et~al.(2024)Xiang, Wu, Zhong, Wagner, Chen, and Mittal}]{xiang2024certifiably}
Chong Xiang, Tong Wu, Zexuan Zhong, David Wagner, Danqi Chen, and Prateek Mittal. 2024.
\newblock Certifiably robust rag against retrieval corruption.
\newblock \emph{arXiv preprint arXiv:2405.15556}.

\bibitem[{Xiong et~al.(2020)Xiong, Xiong, Li, Tang, Liu, Bennett, Ahmed, and Overwijk}]{xiong2020approximate}
Lee Xiong, Chenyan Xiong, Ye~Li, Kwok-Fung Tang, Jialin Liu, Paul Bennett, Junaid Ahmed, and Arnold Overwijk. 2020.
\newblock Approximate nearest neighbor negative contrastive learning for dense text retrieval.
\newblock \emph{arXiv preprint arXiv:2007.00808}.

\bibitem[{Xue et~al.(2024)Xue, Zheng, Hu, Liu, Chen, and Lou}]{xue2024badrag}
Jiaqi Xue, Mengxin Zheng, Yebowen Hu, Fei Liu, Xun Chen, and Qian Lou. 2024.
\newblock Badrag: Identifying vulnerabilities in retrieval augmented generation of large language models.
\newblock \emph{arXiv preprint arXiv:2406.00083}.

\bibitem[{Yang et~al.(2024)Yang, Zhou, Li, and Liu}]{yang2024generalized}
Jingkang Yang, Kaiyang Zhou, Yixuan Li, and Ziwei Liu. 2024.
\newblock Generalized out-of-distribution detection: A survey.
\newblock \emph{International Journal of Computer Vision}, pages 1--28.

\bibitem[{Yang et~al.(2018)Yang, Qi, Zhang, Bengio, Cohen, Salakhutdinov, and Manning}]{yang2018hotpotqa}
Zhilin Yang, Peng Qi, Saizheng Zhang, Yoshua Bengio, William Cohen, Ruslan Salakhutdinov, and Christopher~D Manning. 2018.
\newblock Hotpotqa: A dataset for diverse, explainable multi-hop question answering.
\newblock In \emph{Proceedings of the 2018 Conference on Empirical Methods in Natural Language Processing}, pages 2369--2380.

\bibitem[{Yoran et~al.(2023)Yoran, Wolfson, Ram, and Berant}]{yoran2023making}
Ori Yoran, Tomer Wolfson, Ori Ram, and Jonathan Berant. 2023.
\newblock Making retrieval-augmented language models robust to irrelevant context.
\newblock \emph{arXiv preprint arXiv:2310.01558}.

\bibitem[{Zhong et~al.(2023)Zhong, Huang, Wettig, and Chen}]{zhong2023poisoning}
Zexuan Zhong, Ziqing Huang, Alexander Wettig, and Danqi Chen. 2023.
\newblock Poisoning retrieval corpora by injecting adversarial passages.
\newblock In \emph{2023 Conference on Empirical Methods in Natural Language Processing, EMNLP 2023}, pages 13764--13775.

\bibitem[{Zou et~al.(2024)Zou, Geng, Wang, and Jia}]{zou2024poisonedrag}
Wei Zou, Runpeng Geng, Binghui Wang, and Jinyuan Jia. 2024.
\newblock Poisonedrag: Knowledge corruption attacks to retrieval-augmented generation of large language models.
\newblock \emph{arXiv preprint arXiv:2402.07867}.

\end{thebibliography}

\appendix

\section{Training Details}
\subsection{Dataset} 
\label{sec:train}
\textcolor{blue}{As shown in Table~\ref{tab:example}, we present the average response lengths for both poisoned and correct answers generated by GPT2-XL across three datasets (NQ, HotpotQA, and MS-MARCO), along with examples illustrating each answer format for a specific question. To evaluate the detection of poisoning attacks on the knowledge base of RAG, we selected 3,000 instances of triples $(q, t, a)$ from each of the three evaluation datasets mentioned above. In each triple, $q$ denotes a question, $t$ represents the supporting text collected from Wikipedia or web documents corresponding to $q$, and $a$ is the correct answer to $q$, generated using the state-of-the-art GPT-4 model. Among these 3,000 triplets, 1,500 are randomly selected as benign instances, while the remaining 1,500 are designated as poisoned instances. For each poisoned instance, the poisoned answer $a_p$ is generated by GPT-4 for the given question $q$, and the poisoned text $t_p$ is crafted using existing poisoning strategies, including PoisonedRAG~\cite{zou2024poisonedrag}, GARAG~\cite{cho2024typos}, and PRCAP~\cite{zhong2023poisoning}. The dataset is split into 70\% for training, 20\% for testing, and 10\% as a support set. Within the training set, samples are randomly grouped into triplets (anchor, positive, negative), where the anchor and positive belong to the same class, and the negative belongs to a different class.}

\begin{table*}[t]
    \centering
    \caption{\textcolor{blue}{Statistical data and format of the responses.}}
    \begin{tabular}{p{2.5cm}p{3.5cm}p{8cm}} 
        \toprule
        Dataset & Average Word Count of Response & An Example of Response \\
        \hline
        NQ & Poisoned Response: 7
        
        Correct Response: 12  
        & \textbf{Question}: where is the food stored in a yam plant? 
        
        \textbf{Poisoned Response}: In the leaves. 
        
        \textbf{Correct Response}: In the tuber. \\
        \hline
        HotpotQA & Poisoned Response: 8  
        
        Correct Response: 11  & \textbf{Question}: Which actor starred in Assignment to Kill and passed away in 2000? 
        
        \textbf{Poisoned Response}: Patrick O'Neal. 
        
        \textbf{Correct Response}: John Gielgud. \\
        \hline
        MS-MARCO & Poisoned Response: 16  
        
        Correct Response: 24 & \textbf{Question}: what is hardie plank? 
        
        \textbf{Poisoned Response}: Hardie plank is a wood flooring option that is used for a variety of home styles.
        
        \textbf{Correct Response}: Hardie Plank is a brand of fiber cement siding. \\
        \bottomrule
    \end{tabular}
    \label{tab:example}
\end{table*}

\subsection{Poisoned Texts Generation}
\label{sec:poi}
\textcolor{blue}{To ensure that the retrieved poisoned texts successfully achieve the poisoning effect, we employ three existing methods PoisonedRAG~\cite{zou2024poisonedrag}, GARAG~\cite{cho2024typos}, and PRCAP~\cite{zhong2023poisoning} to generate the poisoned texts. In the PoisonedRAG~\cite{zou2024poisonedrag} method, the attacker first selects a target question along with its corresponding incorrect answer. The attacker then optimizes the design of the poisoned text to ensure that it meets two key criteria: (1) retrievability by the retriever and (2) effectiveness in misleading the language model to generate the incorrect answer. GARAG~\cite{cho2024typos} is a novel adversarial attack algorithm that generates adversarial documents by subtly perturbing clean ones while preserving answer tokens. Through iterative crossover, mutation, and selection, it optimizes the documents to maximize adversarial effectiveness within the defined search space. PRCAP~\cite{zhong2023poisoning} is a gradient-based method, which starts from a natural-language passage and iteratively perturbs it in the discrete token space to maximize its similarity to a set of training queries. }

\textcolor{blue}{It’s worth noting that the generation method for poisoned texts can be any approach that successfully achieves the poisoning effect. Once the activations of both correct and poisoned responses are obtained, we preprocess and use them for training and testing the RevPRAG model. 
}

\subsection{Prompt}
\label{sec:prompt}
\textcolor{blue}{The following is the system prompt for RAG, instructing an LLM to produce a response based on the provided context:}
\begin{figure}[H]
    \centering
    \includegraphics[width=0.99\linewidth]{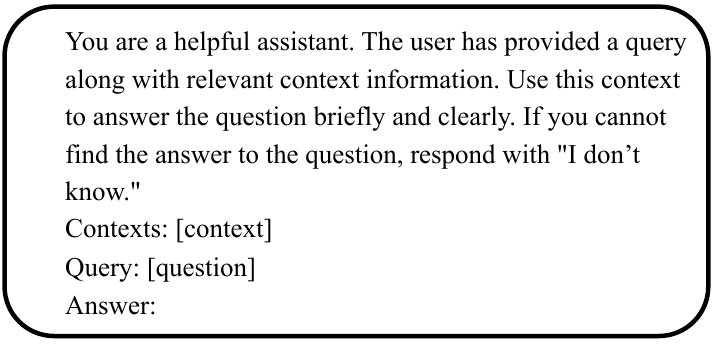}
    \caption{\textcolor{purple}{The prompt used in RAG to make an LLM generate an answer based on the retrieved texts.}}
    \label{fig:prompts}
\end{figure}

\subsection{Environment} 
\textcolor{blue}{We conduct experiments on a server with 64 AMD EPYC 9654 CPUs (64 logical cores enabled) at 2.40–3.70 GHz, 512 GB of DDR5 RAM (assumed based on high-core-count server standards), and four NVIDIA RTX A6000 GPUs, each with 48 GB GDDR6 memory.}

\end{document}